\documentstyle[preprint,aps]{revtex}
\begin{document}
\tightenlines
\title{On the coexistence of diagonal and off-diagonal long-range order,
a Monte Carlo study}
\author{Anne van Otterlo and Karl-Heinz Wagenblast}
\address{Institut f\"ur Theoretische Festk\"orperphysik,
	Universit\"at Karlsruhe, 76128 Karlsruhe, FRG}
\maketitle
\begin{abstract}
The zero temperature properties of interacting 2 dimensional lattice bosons are
investigated. We present Monte Carlo data for soft-core bosons that
demonstrate the existence of a phase in which crystalline long-range order
and off-diagonal long-range order (superfluidity) coexist.
We comment on the difference between hard and soft-core bosons and compare
our data to mean-field results that predict a larger coexistence region.
Furthermore, we determine the critical exponents for the various phase
transitions.
\end{abstract}
\pacs{PACS numbers: 74.20.Mn, 67.90.+z, 05.30.Jp}

The possibility of a phase in dense Bose systems in which diagonal and
off-diagonal long-range order (LRO) coexist has been the subject of discussion
over the past 25 years \cite{kn:m}. Normally bosons at zero temperature are
either superfluid (with off-diagonal LRO) or solid (with diagonal LRO).
However, for a finite range of the interactions between the bosons a
coexistence phase was predicted
within a mean-field approximation \cite{kn:mt,kn:lf,kn:bfs,kn:rs}.
Experiments have been performed on $^{4}He$, but no positive identification of
this coexistence phase (often called supersolid) has yet been made. There are,
however, strong hints towards such a phase \cite{kn:lg}. On the theoretical
side the discussion was restricted to the mean-field level. We are not
aware of any more rigorous studies that identified a supersolid phase.
In this letter we report on Monte Carlo simulations of soft-core lattice
bosons in 2 dimensions that clearly demonstrate the existence of
the supersolid phase beyond the mean-field approximation.

Our lattice boson approach applies especially to artificially fabricated
Josephson junction arrays (JJA) in which the bosons are Cooper pairs that
tunnel between superconducting islands. In these arrays hopping, Coulomb
interactions and the chemical potential may be tuned independently and
therefore they constitute an ideal system
for investigating a supersolid phase. Moreover, the Cooper pairs in
JJA's are soft-core bosons that have a larger coexistence region than
hard-core bosons as was demonstrated
on the mean-field level by Roddick and Stroud \cite{kn:rs}.

The specific model we investigate is defined by the Hamiltonian for a JJA
in the presence of an offset charge $q_{0}$ that corresponds to a chemical
potential for Cooper pairs \cite{kn:bfs}
\begin{equation}
	H= \frac{1}{2}\sum_{ij}(q_{i}-q_{0})U_{ij}(q_{j}-q_{0})-
	E_{J}\sum_{<ij>}\cos(\phi_{i}-\phi_{j}) .
\label{eq:h}
\end{equation}
The number of excess Cooper pairs and the phase of the superconducting order
parameter on island $i$
are denoted by $q_{i}$ and $\phi_{i}$. Number and phase are conjugate
variables that satisfy the commutation relation
$[q_{i},\phi_{j}]=i\delta_{ij}$.
The average density of bosons may be varied by applying a gate
voltage $V_{0}$, which enters in the chemical potential $q_{0}=C_{0}V_{0}/2e$
through the capacitance $C_{0}$ between the islands and the ground
\cite{kn:owfs}. The Coulomb interaction $U_{ij}$ between Cooper pairs
is determined by the inverse of the
matrix that describes the distribution of capacitances between the islands.
We take $U_{ij}$ to be short range, i.e. on-site and nearest neighbor
interactions, $U_{0}$ and $U_{1}$, only. A natural stability condition is
that the number of nearest neighbors
times $U_{1}$ has to be smaller than $U_{0}$. For Cooper
pairs $U_{0}$ is related to the
charging energy $E_{C}$ by $U_{0}=8E_{C}=4e^{2}/C_{0}$.
If the Josephson coupling
energy $E_{J}$ dominates over the Coulomb interaction
$U_{ij}$ the array will be
superconducting at low temperatures. If on the other hand
the Coulomb interaction
dominates a Mott-insulator will be formed \cite{kn:fwgf,kn:owfs}.
The model defined by eq.(\ref{eq:h}) may be mapped onto a Bose-Hubbard model
if the Josephson coupling term is identified
with the hopping-term \cite{kn:fwgf}.

To gain understanding of the zero temperature
properties of the model described by the
Hamiltonian (\ref{eq:h}), we first discuss the
mean-field phase diagram following
Roddick and Stroud \cite{kn:rs}. The phase diagram
is shown in figure 1 a) and b) for
$U_{1}/U_{0}$=0.125 and 0.2 respectively. It
is periodic in $q_{0}$ with period 1 and symmetric
around $q_{0}=\frac{1}{2}$. We discern
four different phases: the superconducting phase (I),
two incompressible Mott-insulating
phases (II and III) and a compressible supersolid phase (IV).
Phases I and IV have a nonzero
superfluid density $\rho_{s}$. Phases III and IV have nontrivial charge order
('checkerboard', see the inset to figure 1) and
therefore a nonzero ($\pi$,$\pi$)-component
of the static structure factor $S_{\pi}$. Thus,
in the supersolid phase LRO ($S_{\pi}\neq 0$)
and off-diagonal LRO ($\rho_{s}\neq 0$) coexist.

The phase diagram for hard-core bosons was investigated
in Refs.\cite{kn:mt,kn:lf,kn:bfs}.
In that limit a supersolid phase is possible only in
the presence of {\it next} nearest
neighbor interactions. The difference with soft-core
bosons is the lack of multiple occupation.
Indeed, the expectation value for 2 soft-core bosons
to be at the same site is nonzero
in phase IV in figure 1 \cite{kn:khw}. We conclude
that the possibility for bosons to hop over
or past each other enhances the supersolid phase.

The points marked $\alpha$, $\beta$ and $\gamma$
in figure 1 have particle-hole symmetry.
This means that the cost in electrostatic energy
is the same for adding or removing a boson.
Point $\alpha$ and the phase boundary between
phases I and II were investigated in
refs.\cite{kn:fwgf,kn:owfs}.
Point $\alpha$ and the line separating phases
I and II have a different dynamical critical
exponent z. This exponent determines the
space-time asymmetry. The correlation length
in the time direction diverges like
$\xi_{\tau}\sim \xi^{z}$, if $\xi$ is the
correlation length in the space directions.
Due to particle-hole symmetry the superconductor-insulator
transition at point $\alpha$ has a dynamical critical
exponent z=1. The transition is in the 3D XY
universality class. For $q_{0}\neq$0 the transition
has z=2 and mean-field exponents apply.
The same holds for point $\beta$ and the line
separating phases III and IV \cite{kn:owfs}.
Motivated by these observations we expect also
point $\gamma$ to have z=1, whereas for the
transition at $q_{0}\neq\frac{1}{2}$ from phase I to IV
we expect z=2. Below we show that this is consistent with
our Monte Carlo data. The points marked $\delta$ in
figure 1 have a first order transition,
as the density jumps from 0 in phase II to $\frac{1}{2}$ in phase III.

Since fluctuations around the mean-field solution are
likely to be important in 2
dimensions one might wonder if the supersolid phase
survives in an exact treatment.
To investigate this question we performed Monte Carlo
simulations of the model
described by the Hamiltonian (\ref{eq:h}).
We follow closely the method used by
S\o rensen et al. \cite{kn:swgy}. Thus, we map
our 2 dimensional quantum model onto
a 3 dimensional classical model of divergence-free
current loops (we use the Villain
form \cite{kn:v} for the cosine in eq.(\ref{eq:h}), see Refs.
\cite{kn:fl,kn:fs} for a derivation). The relevant
quantity is then the partition function
\begin{equation}
	Z=\sum_{\{ J^{\mu}=0,\pm 1, \pm 2, ..\}}
	\exp\left[-\sqrt{\frac{2}{K}} \sum_{ij\tau}J^{\tau}_{i,\tau}
	\left(\delta_{ij}+\frac{U_{1}}{U_{0}}
	\delta_{<ij>}\right)J^{\tau}_{j,\tau}-
	\sqrt{\frac{2}{K}}\sum_{i\tau,a=x,y}
	\left(J^{a}_{i,\tau}\right)^{2} \right] ,
\label{eq:z}
\end{equation}
where the sum is over divergence-free discrete
current configurations that satisfy
$\nabla_{\mu}J^{\mu}=0$ ($\mu$=x,y,$\tau$) and
$\delta_{<ij>}$ equals 1 for nearest neighbors and is
zero otherwise. The time-components of the currents
correspond directly to the particle numbers,
$J^{\tau}_{i}=q_{i}$. The coupling constant
$K=8f E_{J}/U_{0}=f E_{J}/E_{C}$, where $f$ depends on the time-lattice
spacing. Here $f$ is smaller than, but
of the order of unity \cite{kn:khw,kn:v}.

Using the standard Metropolis algorithm
we generate configurations of currents
in a system of size $L\times L\times L_{\tau}$ with
periodic boundary conditions. The generation
of configurations may be done canonical as well
as grand-canonical. Here we work in the
grand-canonical ensemble at fixed $q_{0}$ in
order to make contact to the phase diagrams in figure 1.
As we are interested in a possible supersolid phase,
the relevant quantities to measure are
the superfluid density for off-diagonal LRO and
the structure factor for diagonal LRO.

In terms of the currents $J^{\mu}$ the superfluid density is
\begin{equation}
	\rho_{s}=\frac{1}{L^{2}L_{\tau}}\Big|\sum_{i,\tau}
	J^{x}_{i,\tau} \Big|^{2},
\label{eq:r}
\end{equation}
It satisfies the finite size scaling relation \cite{kn:cfgwy}
$ \rho_{s}=L^{2-d-z}\tilde{\rho}(bL^{1/\nu}\delta, L_{\tau}/L^{z})$
with $\tilde{\rho}$ a
universal scaling function, $b$ a nonuniversal scale factor,
$\nu$ the coherence-length
critical exponent and $\delta=(K-K^{\ast})/K^{\ast}$ the distance to
the transition. At the critical point $K=K^{\ast}$, $\delta=0$
and $L^{z}\rho_{s}$ is a function of
$L_{\tau}/L^{z}$ only. Thus, plots of $L^{z}\rho_{s}$ vs.
$K$ will intersect at the
transition if $L_{\tau}/L^{z}$ is kept constant.
Furthermore, the data for $L^{z}\rho_{s}$
plotted as a function of $L^{1/\nu}\delta$ for
different system sizes should collapse onto
one single curve. This allows the exponent $\nu$ to be determined.

A similar scaling relation holds for the structure factor \cite{kn:nfsb}
\begin{equation}
	S_{\pi}=\frac{1}{L^{4}L_{\tau}}\sum_{ij,\tau}
	(-1)^{i+j} J^{\tau}_{i,\tau}J^{\tau}_{j,\tau} ,
\label{eq:s}
\end{equation}
i.e. $S_{\pi}=L^{-2\beta/\nu}\tilde{S}(b'L^{1/\nu}\delta, L_{\tau}/L^{z})$,
with the order parameter exponent $\beta$.

In the simulations we took $U_{1}/U_{0}$=0.2 in
order to have a large coexistence phase.
We performed simulations for constant $q_{0}$= 0.5,
0.4 and 0 and varied the coupling $K$.
In the phase diagrams in figure 1 this corresponds
to moving on horizontal lines through
the phase transition(s).
For $q_{0}$= 0.5 and 0 we simulated $L \times L \times L$
systems, where $L$= 4, 6, 8, 10, 12, as suggested by particle-hole
symmetry and z=1.
Typically 100,000 sweeps through the lattice were needed for
equilibration and the same
amount for measurement. For $q_{0}$= 0.4 we have z=2. In order
to keep the ratio $L_{\tau}/L^{z}$
constant, we simulated $L \times L \times L^{2}/4$ systems,
where $L$= 6, 8, 10. For the largest
system with $L_{\tau}$= 25 we made up to
400,000 sweeps through the lattice for
equilibration and the double for measurement.
The results are summarized in figures 2-5 and table I.

First we discuss our data for $q_{0}$=0.5. Figure 2 shows that there are two
separate transitions for diagonal and off-diagonal
LRO with a coexistence region in
between where {\it both} the superfluid density
{\it and} the structure factor scale
to a finite value in the thermodynamic limit.
{\it This demonstrates the coexistence
of diagonal LRO and off-diagonal LRO for soft-core
bosons with nearest neighbor
interaction in 2 dimensions.} Fluctuations reduced
considerably the thickness of
the supersolid phase compared to the mean-field
result in figure 1 b. In figures 3 a)
and b) we plot again the same data around the
critical points. In the neighborhood
of the critical points the data fall onto a
single curve when plotted as a function
of $L^{1/\nu}\delta$. Table I shows that the
exponent $\nu$ is different for the two
transitions. For the transition related to
superfluidity (point $\beta$ in figure 1) we
find a value for $\nu$ that is consistent
with the 3D XY universality class which has
$\nu\approx$0.67. For the transition related
to crystalline order (point $\gamma$) we
find $\nu\approx$0.55 and $\beta\approx$0.21.

Also at $q_{0}$=0.4 we find two separate transitions
that are the boundaries for
the supersolid phase in between, see figure 4 and
table I. As compared to $q_{0}$=0.5
both transitions are shifted to smaller values of
the coupling constant $K$. This
is consistent with the mean-field phase diagram.
Again the two transitions have
different critical exponents. The transition related
to superfluidity (the line
separating phases III and IV in figure 1) has $\nu\approx$0.5
which is consistent with a mean-field transition in d+z=4
effective dimensions.
The transition related to crystalline order
(between phases I and IV) has an order-parameter
exponent $\beta\approx$0.25. This rules out
a mean-field transition for diagonal LRO,
although the transition is effectively 4-dimensional.
In the neighborhood of this
transition, fluctuations of the x,y-components of
the currents $J$ induce long-range
interactions in the time direction for the
$\tau$-components of the currents $J$ \cite{kn:khw}.
It is likely that these long-range interactions
are a relevant perturbation and suppress the
exponent $\beta$.

Finally the data for $q_{0}$=0 are shown in figure 5.
Here there is only one phase
transition, as the Mott-insulating lobes (phase II in
figure 1) do not have any non-trivial
crystalline order. Our data are consistent with a
transition in the 3D XY universality class.

In conclusion we have performed Monte Carlo simulations
on soft-core lattice bosons
in two dimensions that establish the existence of a
supersolid phase in which diagonal
and off-diagonal long-range order coexist. We estimated
critical exponents as listed in table I.
The mean-field phase diagram of ref.\cite{kn:rs}
is qualitatively confirmed. However, our simulations
indicate that the supersolid phase
is smaller than one would deduce from mean-field theory.
We suggest that the coexistence
phase may be observed in two-dimensional systems such as
Josephson junction arrays or
thin $^{4}He$-films on suitable substrates. In these
systems the possibility to vary
the coupling constants as well as the chemical potential
should make it possible to
tune through the supersolid phase and see two sequential phase transitions.

We acknowledge the help and suggestions of R. Fazio,
C. Bruder, G. Sch\"{o}n and G.T. Zimanyi.
This work is part of ``Sonderforschungsbereich 195''
which is supported by the Deutsche
Forschungsgemeinschaft.

\vspace{1cm}

\begin{table}
$$
\begin{array}{|c|c|c|}
\hline\hline
\multicolumn{1}{|c|}{} &\multicolumn{2}{c|}{\mbox{the
transition for}} \\ \cline{2-3}
\hspace{25pt}q_{0}\hspace{25pt}       &\hspace{40pt}
\mbox{off-diagonal LRO}  \hspace{40pt}
      &  \hspace{50pt}\mbox{diagonal LRO}
\hspace{50pt}    \\ \hline \hline
0.5           & K^{\ast}=0.775\pm 0.005
& K^{\ast}=0.837\pm 0.005\\ \cline{2-3}
(z=1)         & \nu=0.65\pm 0.08
& \nu=0.55\pm 0.05, \beta=0.21\pm 0.04 \\ \hline \hline
0.4           & K^{\ast}=0.645\pm 0.008
& K^{\ast}=0.749\pm 0.006 \\ \cline{2-3}
(z=2)         & \nu=0.44\pm 0.08
& \nu=0.5\pm 0.1, \beta=0.25\pm 0.10  \\ \hline \hline
0.0           & K^{\ast}=0.843\pm 0.005
& \mbox{-----------} \\ \cline{2-3}
(z=1)         & \nu=0.61\pm 0.08
& \mbox{-----------} \\ \hline\hline
\end{array}
$$
\caption{Critical couplings and exponents for the different transitions}
\label{table}
\end{table}

\vspace{1cm}

\begin{large}
\noindent
Figure Captions :
\end{large}

\vspace{1cm}

\noindent
Fig. 1 : Phase diagrams for soft-core bosons with
on-site and nearest neighbor interaction.
a) $U_{1}/U_{0}$=0.125. b) $U_{1}/U_{0}$=0.2.
There are 4 different phases, I: superconductor,
II: Mott-insulating, III: Mott-insulating with
checkerboard charge-order and IV: supersolid.
The points marked $\alpha$, $\beta$ and $\gamma$
have particle-hole symmetry. The points marked
$\delta$ have a first order transition. The inset
to a) shows the checkerboard charge order.

\vspace{1cm}

\noindent
Fig. 2 : Data for $L\rho_{s}$ and
$L^{\frac{2\beta}{\nu}}S_{\pi}$ with $\frac{2\beta}{\nu}$=0.78
vs. $K$ at $q_{0}$=0.5. The curves cross at
$K^{\ast}$=0.775 and 0.837 respectively.
The region in between is the supersolid phase.

\vspace{1cm}

\noindent
Fig. 3 : Data for $\rho_{s}$ and $S_{\pi}$ at
$q_{0}$=0.5 in the neighborhood of the two
critical points, scaled as to collapse onto a
single curve. The drawn lines are a low order
polynomial fit to the data. a) $L\rho_{s}$ vs. $\delta L^{1/\nu}$
with $\nu$=0.65. b) $L^{0.78}S_{\pi}$ vs. $\delta L^{1/\nu}$ with $\nu$=0.55.

\vspace{1cm}

\noindent
Fig. 4 : Data for $\rho_{s}$ and $S_{\pi}$ at $q_{0}$=0.4.
The drawn lines are a low order
polynomial fit to the data. a) $L^{2}\rho_{s}$ vs. $K$.
The curves cross at $K^{\ast}$=0.645. b)
$L^{\frac{2\beta}{\nu}}S_{\pi}$ vs. $K$ with
$\frac{2\beta}{\nu}$=1.0. The curves cross at $K^{\ast}$=0.749.

\vspace{1cm}

\noindent
Fig. 5 : Data for $\rho_{s}$ at $q_{0}$=0. The drawn lines are a low order
polynomial fit to the data. $L\rho_{s}$ vs. $K$. The curves cross
at $K^{\ast}$=0.843.

\end{document}